\documentclass[12pt]{elsarticle}
\usepackage{lineno,hyperref}
\usepackage[latin1]{inputenc}
\modulolinenumbers[5]

\journal{SHPMP}

\let\today\relax
\makeatletter
\def\ps@pprintTitle{%
    \let\@oddhead\@empty
    \let\@evenhead\@empty
    \def\@oddfoot{\footnotesize\itshape
         {Forthcoming in Studies in History and Philosophy of Modern Physics} \hfill\today}%
    \let\@evenfoot\@oddfoot
    }
\makeatother

\makeatother

\hypersetup{                 
    colorlinks=true,         
    linkcolor=blue,          
    citecolor=blue,           
    urlcolor=blue          
}
\begin{document}

\begin{frontmatter}

\author{Ana-Maria Cre\c{t}u\fnref{myfootnote}}
\address{School of Philosophy, Psychology and Language Sciences, University of Edinburgh}
\fntext[myfootnote]{d.cretuanamaria@gmail.com}

\title{Diagnosing Disagreements:\\ The Authentication of the Positron 1931 - 1934}

\begin{abstract}
This paper bridges a historiographical gap in accounts of the prediction and discovery of the positron by combining three ingredients. First, the prediction and discovery of the positron are situated in the broader context of a period of `crystallisation' of a research tradition. Second, the prediction and discovery of the positron are discussed in the context of the `authentication' of the particle. Third, the attitude of the relevant scientists to both prediction and discovery are conceptualised in terms of the idea of `perspectives'. It will be argued that by examining the prediction and discovery of the positron in the context of authentication within a period of crystallisation, we can better understand disagreements regarding the positron between relevant scientists (Dirac, Bohr, and Pauli) in the period 1931-34.
\end{abstract}


\end{frontmatter}

\newpage
\section{Introduction} 
The Solvay Conference of 1933 marks the recognition of the positron as a new particle of matter. However, the discussions during the conference brought out prior disagreements with regard to the nature of the positron. Whilst the historical details of the prediction and discovery of the positron have already been subject to a reasonably complete existing literature \cite{Hanson:1961,Darrigol:1988,Kragh:1990,Roque:1997,Pashby:2012,Leone:2012}, important questions related to the disagreements attending the nature of the positron remain nevertheless puzzling. In particular: Why were Bohr and Pauli so opposed to equating the experimental particle with the Dirac-particle? Were their seemingly mysterious attitudes born of dogmatic conservatism? What factors led to their eventual change of heart? 

This paper bridges a historiographical gap by offering a novel combination of three ingredients. First, the prediction and discovery of the positron are situated in the broader context of a period of `crystallisation' of a research tradition. Second, the prediction and discovery of the positron are discussed in the context of the `authentication' of the particle. Third, the epistemic attitudes of the relevant scientists are conceptualised in terms of the idea of `perspectives'. Crystallisation leads physicists to reevaluate core assumptions. In this context, scientists with different epistemic perspectives may respond differentially to the same empirical knowledge, this in turn may serve to impede any process of authentication. In this paper it will be shown that examining the prediction and discovery of the positron in the context of authentication within a period crystallisation can help explain the mystifying attitudes of certain physicists both prior and throughout the 7th Solvay Conference, particularly the epistemic attitudes of Bohr and Pauli. 

The remainder of the paper is structured as follows. In Section 2 the necessary conceptual background is briefly summarised. Section 3 gives an overview of the relevant historical context of the crystallisation of a research tradition. Section 4 details Dirac's prediction and the disagreements that ensued. In section 5 the perspectives of Pauli and Bohr are evaluated, and in Section 6 the process of the authentication of the positron is considered. 

\section{Conceptual Background}

Following Laudan \cite{Laudan:1977}, a research tradition can be understood as ``the set of assumptions ... about the basic kinds of entities in the world, assumptions about the proper methods to use for constructing and testing theories about those entities" (p. 97), in brief, ``a set of ontological and methodological `do's' and `don'ts'"(p. 80). Research traditions foster a variety of different theories which normally inherit the set of assumptions of their parent research tradition. Crystallisation marks the early stages of a research tradition in which a wealth of \textit{conceptual and empirical problems} are still in need of resolution. Empirical problems concern empirical phenomena in need of explanation, whilst conceptual problems are due to inconsistencies between the theories fostered by the research tradition. The paradigmatic example of crystallisation is quantum theory in the early 20th century (1900 -1925), see for example \cite{Kao:2015}. Authentication then refers to the validation of a particular empirical phenomena as a well established, genuine, empirical effect. Unlike Laudan, in this paper authentication is not restricted to the validation of experimental results, of which, in the case of the positron there already exists a vast literature (\cite{Roque:1997,Pashby:2012,Leone:2012}, etc.). Whilst the experimental authentication of the positron is instrumental to the confirmation of the positron as a new particle of matter, the focus here is on the theoretical authentication which was equally instrumental to its proper recognition. 

Physicists' epistemic attitudes can be described in terms of `perspectives'. \textit{A perspective can be understood as a sophisticated theoretical framework that encompasses the set of theoretical interests and background theoretical knowledge (principles and assumptions equally) that a researcher or group of researchers can be said to hold at any given time.} A perspective represents the working stance of a scientist. Perspectives are conceptual bookkeeping tools for tracking relevant epistemic shifts in scientists' attitudes towards a given conceptual or empirical problem. The notion of perspective, as used here, does not encompass personal, idiosyncratic likes or dislikes of the relevant scientists. Thus perspectives, insofar as they can be reconstructed from primary and secondary historical sources, reflect the epistemic or theoretical attitude of a scientist, and not their subjective, idiosyncratic attitude towards a given problem. Tracking a scientist's perspective can be particularly useful in cases where the solution to a particular conceptual or empirical problem is under discussion and disagreement ensues. In particular, tracking any relevant shifts in a scientist's perspective may help one understand the nature of the disagreement and thus lead to its resolution. Similarly to Laudan's theories, perspectives inherit the theoretical assumptions of the research tradition. It is further worth noting that the notion of perspectives used here is different from the notion of perspective as defined by perspectival realists such as Massimi or Giere. Massimi \cite{Massimi:2016a,Massimi:2018a} takes perspectives to encompass both ``(i) the body of scientific knowledge claims advanced by a scientific community at the time; (ii) the experimental, theoretical, and technological resources available to the scientific community at the time to reliably make those scientific knowledge claims; and (iii) second-order (methodological-epistemic) claims that justify the scientific knowledge claims advanced'', and thus her notion of `perspective' is closer to Laudan's research tradition. In contrast to Massimi's notion, `perspective' as defined here is more localised, yet not as localised as Giere's (\cite{Giere:2006,Giere:2010}) perspectives. For Giere, a perspective is a `theoretical' or `observational' viewpoint afforded by a theory, whereas on the view advanced here perspectives can encompass elements of different theories at the same time. 

In periods of crystallisation, perspectives can deviate from the core assumptions of a research tradition and can affect the authentication of certain phenomena. Crystallisation, being characterised by both unsolved conceptual and empirical problems, and sometimes, by rapid discoveries, can deepen and diversify the reevaluation of core assumptions according to the theoretical interests of a particular scientist. Thus, scientists with different perspectives may respond differentially to the same empirical knowledge, hence impeding authentication. In the remainder of this paper it will be argued that once examined against the background of crystallisation, the perspectives of Bohr and Pauli towards the authentication of the positron are no longer puzzling.

\section{Crystallisation}
The basic argument put forward in this paper is that crystallisation leads physicists to reevaluate core assumptions. In this context, scientists with different perspectives may respond differentially to the same empirical knowledge, and this in turn may serve to impede any process of authentication. Furthermore, disputes regarding authentication within crystallisation can lead to productive forms of disagreement.\footnote{ Here, we follow \cite{DeCruz:2013}, who introduce the idea of `productive disagreement' to describe situations in which, in the absence of new decisive evidence, scientists are better off maintaining disagreement, as ``maintaining epistemic peer disagreement is valuable for furthering epistemic goals" (p. 170).} Once understood against the background of crystallisation, the puzzling perspectives of scientists in regard to the authentication of an empirical phenomena stop having a mystifying character. The first step in establishing the basic argument of this paper is to put forward evidence of crystallisation. Thus, this section will provide a brief account of the main assumptions and of the main conceptual and empirical problems of the early 1930s.

Up until the discovery of the neutron in 1932,\footnote{ Chadwick's letter to Nature announcing the discovery of the neutron was published on February 27 1932, see \cite{Chadwick:1932}. For historical accounts of the neutron discovery and ensuing controversies see \cite{Stuewer:2012}, and \cite{Bromberg:1971}.} most physicists thought that there exist only two kinds of fundamental particles: the proton and the electron. As a consequence of this, most physicists believed that the atomic nucleus was made of protons and electrons (see \cite{Stuewer:2012,Stuewer:2018}). However, the confinement of electrons inside the nucleus led to a wealth of problems concerning the stability of nuclei, their spin, and their statistics.\footnote{ For a description of the nuclear paradoxes see \cite{Pais:1991}, pp. 330-332.} In addition to troubles  with the structure of the nucleus, the continuous spectrum of  $\beta$-decay was still in need of explanation.\footnote{ For details regarding the $\beta$-decay controversy see \cite{Guerra:2014}, \cite{Stuewer:2012}, \cite{Jensen:2000}, \cite{Brown:1988}, and \cite{Bromberg:1971}.} Some of these difficulties seemed so great that a breakdown of quantum mechanics at the nuclear level was not considered out of the question (see \cite{Darrigol:1988}, p. 228). Thus before the discovery of new particles, physicists struggled to apply the new quantum mechanics to nuclear phenomena and physicists such as Bohr, for example, called into question core assumptions of the research tradition (e.g. the energy conservation principle).\footnote{ Bohr's `attack' of the energy conservation principle is well known and has been discussed in a variety of papers, most recently by \cite{Kragh:2017}and \cite{Guerra:2014}; it will also be the subject of more detailed discussion in what follows.} Whilst in 1932 new particles were discovered, these new discoveries did not immediately lead to solutions to the aforementioned problems.\footnote{ For a review of the `new particles' in the 1930, see \cite{Stuewer:2018}, ch. 7, \cite{Guerra:2018}, ch. 5, and \cite{Navarro:2004}.} Some of the particles were accepted more readily than others, for example the neutron,\footnote{ This is not to say that the authentication of the neutron proceeded in an entirely uncontroversial manner, but whilst its nature was debated, its existence was accepted almost immediately.} whilst in other cases, such as the positron's, their discovery marked the beginning of a long process of authentication. It is against this background -- which can be understood as a period of crystallisation of a research tradition -- that the prediction of the positron was made and its authentication begun. 

A major turn in relativistic quantum mechanics took place with Dirac's derivation of his relativistic equation of the electron (\cite{Dirac:1928}).\footnote{ For Dirac's own recollections on the origin of the equation see \cite{Dirac:1980,Dirac:1973,Dirac:1962}. For reviews on the origins of Dirac's equation see \cite{Moyer:1981a} and \cite{Kragh:1981}.} A range of spectroscopic anomalies, such as the Zeeman effect, were explained by introducing spin\footnote{ For an early history of the Spin-Statistics Theorem see \cite{Massimi:2005} and \cite{Blum:2014}.}, yet neither spin nor relativity were yet integrated in quantum mechanics (\cite{Kragh:1981}, p. 44). With Dirac's equation, ``spin was a necessary consequence, the correct magnetic moment of the electron was obtained, the anomalous Zeeman effect of atoms came out right, the Thomas factor of the electron appeared automatically" (\cite{Pais:1989}, p. 95). Despite these spectacular results, Dirac's theory of the electron had also some undesirable consequences.\footnote{ For example, Moyer \cite{Moyer:1981b} notes that `` Two metaphors -- magic and sickness -- recur in recollections of early reactions to Dirac's theory. Oppenheimer, for example, used both in close juxtaposition. The theory magically yielded properties of the electron from general formal considerations. But the theory was also afflicted with the sickness of predicting properties not observed" (p. 1056).} The equation had four solutions: two of the solutions referred to \textit{free} electrons, i.e., electrons with negative charge, and two solutions referred to \textit{electrons with negative energy}, i.e., electrons with positive charge -- which could not at the time be given any physical interpretation.\footnote{ ``Only half of the solutions to the Dirac equation for an interacting electron refer to the particle's negative charge; the other half can be shown to refer to a particle of positive charge" (\cite{Kragh:1990}, p. 88).} Whilst the negative energy solutions were not a new theoretical concept, as they beset the Klein-Gordon equation as well, the negative energy problem had not been carefully studied in relation to the Klein-Gordon equation because this equation was considered theoretically problematic (it had no one-particle interpretation) and empirically inadequate (it did not give the hydrogen fine structure).\footnote{ Thanks to Alexander Blum for pointing out some of the problems related to the Klein-Gordon equation.} But since the Dirac equation gave such spectacular results, the negative solutions couldn't be ignored, at least not indefinitely. 

Dirac himself initially chose to ignore the negative solutions,\footnote{ ``Since half the solutions must be rejected as referring to the charge + e on the electron, the correct number will be left to account for duplexity phenomena" \cite{Dirac:1928}, p. 618.} but in light of the Klein paradox and the Klein-Nishina formula,\footnote{ Klein's paradox showed that ``in some simple cases the Dirac electron would behave in a patently absurd way" (\cite{Kragh:1990}, p. 89), whilst the Klein-Nishina scattering formula ``relied crucially on positive as well as negative energies" (\cite{Kragh:1990}, p. 89).} some explanation of  the difficulty of electrons having negative kinetic energy (henceforth the `$\pm$ difficulty') had to be found. However, since the negative energy solutions were not believed to be a real phenomena, there was no easy way to provide a satisfactory explanation to the `$\pm$ difficulty'. Thus whilst not being an immediate consequence of Dirac's relativistic equation of the electron, it is the `$\pm$ difficulty' that led, by degrees, to the prediction of the positron.\footnote{ Contrary to a naive account, the positron is not an immediate consequence of Dirac's equation. In fact, Pashby notes that ``The \textit{presence} of negative energy solutions was not enough to predict anti-matter: each of the negative energy states had to be occupied by an electron for `anti-electrons' (holes) to have physically reasonable properties, and moreover properties which in detail turned out to be instantiated by the positron" (p. 462).}

Dirac's equation, though providing a range of excellent results, added to the problem of crystallisation. The `$\pm$ difficulty' and a bit later the `hole theory' called into question core symmetry and conservation assumptions, thus exacerbating the need for reevaluation. In the next section it will therefore prove instructive to trace the sequence of events that led Dirac from initially dismissing the `$\pm$ difficulty' to the `hole' theory for at least two reasons. First, because Dirac's theoretical prediction of the positron marks the first step in its authentication and also leads to further reevaluation of certain core assumptions. Second, because the prediction of the positron and reactions thereto constitute further evidence of crystallisation and of its effects on authentication. 

\section{Pre-Authentication}
Dirac's first attempt to solve the ``$\pm$ difficulty" was not only an attempt to overcome the Klein paradox; it was also prompted by a letter from Niels Bohr who foreshadowed more stringent problems arising from the application of Dirac's equation to ``nuclear electrons''.\footnote{ See \cite{Pashby:2012}, \cite{Kragh:1990}, and \cite{Darrigol:1988} for details.} Bohr alluded to the possibility that the ``$\pm$ difficulty" ``might be said to reveal a contrast between the claims of conservation of energy and momentum on one side and the conservation of the individual particles on the other side" (\cite{BCW:1986}, p. 547). Dirac's response took the form of his first articulation of the `hole' theory. The hole theory assumes that there is an infinite sea of negative electrons, in which ``most stable states are occupied [...] except perhaps a few of small velocity'' (\cite{Dirac:1930a}, p. 362). The unoccupied states or the \textit{holes} were initially identified by Dirac with protons \cite{BCW:1986}, p. 549). Herman Weyl, also, independently showed that the negative energy solutions could be explained away by interpreting ``the four differential equations as including the proton in addition to the electron" (\cite{Weyl:1950}, p. 225), only to later dismiss this idea and conclusively show that holes could not be protons. Weyl showed that the mass of electrons and holes is the same\footnote{ See \cite{Pashby:2012} for more details. See also \cite{Weyl:1950}.}, a conclusion supported also by Tamm and Oppenheimer (\cite{Pais:1987}, p. 101). However, in between the two stages of Weyl's proofs, Dirac showed why the immediate identification of a `hole' with the proton \`{a} la Weyl (i.e., the early proof) was afflicted by a number of problems which Dirac's (\cite{Dirac:1930a, Dirac:1930b}) more elevated proof of why the `holes' should be identified with the protons seemed to eschew. 

The straightforward assertion ``that a negative-energy electron \textit{is} a proton" generated, according to Dirac, paradoxes related to the conservation of electric charge, conservation of momentum, and concerns with regard to the physical reality of a negative electron (\cite{Dirac:1930a, Dirac:1930b}). The problematic nature of the negative-energy electron, and hence its immediate identification with the proton, consisted in the fact that it would have ``less energy the faster it moves and it will have to absorb energy in order to be brought to rest'' (\cite{Dirac:1930a}, p. 362) and it would be difficult to understand how such a particle could ever be observed in nature. Finding a way around these difficulties whilst accounting for the conditions in which the ``unwanted solutions with negative kinetic energy for the electron"\footnote{ See \cite{Dirac:1930a}, p. 360.} could hold in the actual world, led Dirac to further articulate his `hole theory' and ingeniously explain why it would seem that the proton is the ideal candidate for a hole. Dirac subsequently thought that the hole theory went beyond the purely mathematical considerations that led Weyl to associate the negative energy electrons with the protons. The hole theory showed how ``the states corresponding to a negative energy" can have ``some physical meaning"(\cite{Dirac:1930b}). Based on the assumption that all the states of negative energy for an electron are occupied, together with the fact that ``an electron with negative energy moves in an electromagnetic field in the same way as an ordinary electron with positive energy would move if its charge were reversed in sign, so as to be +\textit{e} instead of -\textit{e}" (\cite{Dirac:1930b}), as well as consideration of Pauli's exclusion principle, Dirac was led to ``assert that \textit{the hole is a proton}"(\cite{Dirac:1930b}). 

The identification of `holes' with protons did not, however, alleviate many, if any, of the initial concerns regarding the ``$\pm$ difficulty". On the contrary, it immediately led to further difficulties (e.g. symmetry violations \footnote{ ``[...] the Dirac wave equation is symmetric with respect to negative and positive charges (electrons and anti-electrons), while nature shows no symmetry between the electron and the much heavier proton" (\cite{Kragh:1990}, p. 96).}) which brought about an increasing dislike of the hole theory. In fact:

\begin{quote}
In its first two years Dirac's hole theory was certainly not as popular as Dirac's equation. It was flatly rejected by Bohr, who maintained his view that the negative energy difficulty announced a breakdown of fundamental concepts, and, in spite of a short period of interest, it was mocked by Pauli (\cite{Darrigol:1988}, p. 238).
\end{quote}

The difficulties introduced by the proton hypothesis were mainly due to the mass asymmetry between the electron and the proton. The mass asymmetry, however, was not the only problem introduced by the hole theory. The identification of `holes' with protons would also lead to ``paradoxes regarding conservation of charge, momentum, and energy" (\cite{Moyer:1981b}, p. 1058) as well as potential violations to the known stability of matter. Moreover, considerations regarding the physical meaning of `spin' would lead some (e.g. Bohr) to believe that the negative-energy solutions show a breakdown of various concepts at the quantum-relativistic level. In particular, Bohr was worried that the ``property called `spin' could not be interpreted because there was no corresponding classical property" (\cite{Moyer:1981b}, p. 1058). Fortunately, the `proton' hypothesis was short-lived: in 1931 Dirac identified the hole with a new kind of particle that should have the same mass as the electron but positive charge: the anti-electron (i.e., the positron): 

\begin{quote}
A hole, if there were one, would be a new kind of particle, unknown to experimental physics, having the same mass and opposite charge to an electron. We may call such a particle an anti-electron. We would not expect to find any of them in nature, on account of their rapid rate of recombination with electrons, but if they could be produced experimentally in high vacuum they would be quite stable and amenable to observation. An encounter between two hard $\gamma$-rays (of energy at least half a million volts) could lead to the creation simultaneously of an electron and anti-electron, the probability of occurrence of this process being of the same order of magnitude as that of the collision of two $\gamma$-rays on the assumption that they are spheres of the same size as classical electrons. This probability is negligible, however, with the intensities of $\gamma$-rays at present available. The protons on the above view are quite unconnected with electrons. Presumably the protons will have their own negative-energy states, all of which normally are occupied, an unoccupied one appearing as an anti-proton. Theory at present is quite unable to suggest a reason why there should be any differences between electrons and protons (\cite{Dirac:1931}, pp. 61 -- 62).
\end{quote}

Dirac's theoretical prediction of the anti-electron is unanimously recognised by physicists, historians of science, and philosophers (\cite{Anderson:1961}, \cite{Hanson:1961}, \cite{Pais:1998}, \cite{Jacob:1998}, \cite{Griffiths:2004}, \cite{Pashby:2012}) as the \textit{bone fide} prediction of the positron. However, at the time of the prediction, no experiments were yet conducted to search for the anti-electron or positron, and Dirac's theory was neither well received, nor was the association between the subsequent experimental results immediately associated with it. In fact, it would take a long time for Dirac's prediction to be taken seriously. For example, Heilbron recollects of his interview with P.M.S. Blackett that ``[w]hen asked how long they had known about Dirac's theory, Blackett replied he wasn't quite certain, but that it didn't matter anyway because nobody took Dirac's theory seriously'' (\cite{Heilbron:1962}). Also, Chadwick, in an interview with Wiener, recollects that ``not a great deal of notice was taken at once about this hole or positive electron -- not a great deal, not until a year or two later" (\cite{Chadwick:1969}). Though it is hardly controversial to claim that the authentication of the positron started with its theoretical prediction, recollections such as Chadwick's and Blackett's suggest that the theoretical aspect of the authentication hindered, rather than helped the recognition of the positron as a new particle of matter. Even when Anderson's experiments were put forward, few if any associated the experimental evidence with Dirac's prediction.\footnote{ Rosenfeld, in an interview with Oskar Klein, notes that ``[there was a very strong resistance in Copenhagen, from Bohr, when the announcement of the positive electron came. Bohr would not believe it; he would not believe it. He did not make, and nobody here made the connection with Dirac's theory. I think somebody may have mentioned Dirac theory, and then Bohr dismissed it by saying, `Oh, but that is no theory. That is not a theory that one can believe in'" \cite{Klein:1963}.} It wasn't until Blackett and Occhialini's experimental results were put forward that Dirac's theoretical prediction started to be taken more seriously. However, Blackett and Occhialini's experiments led neither to the immediate recognition of the positron as a new particle of matter nor to the confirmation of Dirac's theory. Their experiments, though received with enthusiasm by some,\footnote{ ``I got used to say, that your prediction of the antielectron has no parallel in the history of science. [...] Your theoretical prediction about the existence of the antielectron, being unstable in the `ordinary space' outside the nucleus, seemed so extravagant and \textit{totally} new, that you yourself dared not cling to it and preferred rather to abandon the theory. And now the experiment unexpectedly proved you to be right ..." (\cite{Kojevnikov:1993}, p. 64).} led others to express a deep opposition towards the experimental particle, and an even stronger opposition to the equation of the experimental particle with the Dirac particle. 

Though Dirac himself has suggested that experimentalists did not see the `positron' due to `prejudice', and that it was also `prejudice' that led Bohr, Pauli, and others to oppose the positron,\footnote{ ``Many of the developments which have been occurring in recent times have been concerned with the introduction of new particles. There again we see that the physicist had to overcome a prejudice. Up until about 1930 physicists thought that there were only two fundamental particles, the electron and the proton. [...] There was very strong reluctance to postulate new particles up until about that time. [...] Why did experimentalists not see them? Because they were prejudiced against them" (\cite{Dirac:1973},  pp. 12 -13).} there are reasons to resist this explanation. Bohr and Pauli were not completely opposed to new particles since they already accepted the neutron as a new particle of matter.\footnote{ Note that though the neutron's discovery did prompt a certain amount of disagreement regarding its nature, the recognition of the neutron as a genuine empirical phenomena, i.e., its authentication, preceded such disagreement.} Moreover, Bohr was ready to part with the laws of conservation, an attitude anything but conservative; whilst Pauli himself entertained the possibility of the existence of a new particle, i.e., the neutrino, again an attitude that does not exude prejudice. Thus explaining the opposition of Bohr and Pauli towards the positron in terms of dogmatic conservatism, cannot be the whole story. It is more likely that due to the effects of crystallisation Pauli's and Bohr's opposition to the authentication of the positron ought to be sought in the effect of the positron on their working perspectives. 

It has so far been shown that the prediction and discovery of the positron took place against the background of crystallisation. Despite three separate events suggesting the existence of the positron -- i.e., Dirac's theoretical prediction, Anderson's discovery, Blackett' and Occhialini's confirmatory experiments -- this new particle was not fully authenticated till the 7th Solvay Conference of 1933. Whilst the existing historical and philosophical literature provides a good deal of evidence for the existence of disagreements in connection to the confirmation of the positron as a new particle of matter, a definitive diagnosis of the \textit{causes} of these disagreements has yet to be established. 

Hanson, for example, argues that ``the discovery of the positive electron was a discovery of three different particles" (\cite{Hanson:1961}, p. 194). Whilst his analysis provides insight into The Anderson Particle, The Dirac Particle, and The Blackett-Occhialini Particle, Hanson does not offer any definitive pronouncement with regard to what led to the equation of the Dirac particle with the experimental particle. In addition to Hanson's three particles, Darrigol \cite{Darrigol:1988}, and in passing Roqu\'e \cite{Roque:1997}, draw attention to a fourth particle with positive charge, mass similar to the electron's, but of \textit{integer spin}, i.e., the Bose-positron. This was an interpretative hypothesis emerging on the back of Anderson's experiments, which albeit short lived was for some time regarded as a possible solution to some of the conceptual and empirical problems of the early 1930's. Darrigol draws attention to the disagreement regarding the interpretation of the experimental results and highlights the fact that a number of physicists were reluctant to equate the experimentally discovered particle (i.e., Anderson's positive electron) with the theoretically `discovered' particle (i.e., Dirac's anti-electron).\footnote{ See Darrigol's quote on p. 5.} 

Roqu\'e, in an insightful account of how the positron was established as a new particle, calls attention to a deeper disagreement regarding the status of cosmic rays experiments, which were used to obtain the famous photographs of positron tracks published by Anderson in 1932. According to Roqu\'e, serious doubt was entertained with regard to ``the evidential weight of experiments involving cosmic rays" (\cite{Roque:1997}, p. 85), urging researchers to find means of producing ``positive electrons by means of radioactive sources, rather than undependable cosmic rays" (ibid., p. 88). Roqu\'e thus argues that it was the successful `manufacture' that resolved the disagreements.

Finally, Pashby \cite{Pashby:2012}, in an incisive paper on Dirac's prediction of the positron, provides additional reasons for concern regarding the debate over the theoretical identity of the positron. Whilst forcefully arguing that Dirac's prediction of the positron ought to be seen as a \textit{bone fide} theoretical prediction, and hardly anyone would deny this, he also points out that ``in the conceptual framework of the hole theory, there could be no antimatter without the Dirac sea' (\cite{Pashby:2012}, p. 462), and hence no positron either. Pashby thus emphasises the importance of the theoretical side of the debate regarding the identity of the positron. 
 
The existing literature notwithstanding, a diagnosis of the disagreements has yet to be established. What we suggest is that the disagreements regarding the positron are less mystifying if they are understood in the context of authentication within crystallisation. Crystallisation leads physicists to reevaluate core assumptions and this reevaluation, in turn, can impede authentication. Thus whilst Roqu\'e  is right to suggest that the slow authentication of the positron can be attributed to the need of ``manufacturing positrons'', this can only be a partial diagnosis. To account for the theoretical disagreements, the resistance on the part of leading physicists such as Pauli and Bohr to the identification of the experimental particle with the Dirac particle must be accounted for. In the following section, it will be argued that their attitudes can be shown to be neither dogmatic nor mysterious, provided that they are understood in the context of authentication within crystallisation.

\section{Perspectives}
Throughout 1933 leading physicists such as Bohr and Pauli were sceptical of the experiments involving positrons and opposed the identification of the experimental particle with Dirac's theoretical prediction. Barring dogmatic conservatism, their attitudes are strikingly puzzling. Why did they oppose so vehemently and for such a long time the identification of the experimental particle with Dirac's prediction? This section is dedicated to an examination of Pauli's and Bohr's respective perspectives from which a richer explanation of their attitudes towards the positron is to be established. However, before we examine in more detail Bohr's and Pauli' perspectives, a few clarifications on perspectives themselves are in order. 

It is useful to examine a scientists' epistemic attitude towards a particular conceptual or empirical problem by first understanding their `perspective'. The perspective of a scientist encompasses both their theoretical interests and their background theoretical knowledge. First, in regard to interests, it is worth emphasising that scientists can, at any given point, apply themselves to a number of different conceptual or empirical problems. Second, in regard to background theoretical knowledge, it is useful to note that in attempting to resolve any conceptual or empirical problem, scientists will have at their disposal the background assumptions of the research tradition, as well as less entrenched and less general assumptions pertaining to the relevant solutions they're working on in relation to the problems they apply themselves to. It is worth reemphasising that the notion of perspective used here is an epistemic one, referring to a scientists's theoretical stance and not to their subjective, or idiosyncratic attitude. A scientists' perspective, particularly in periods of crystallisation, may lead them to respond differentially to the same empirical knowledge. By examining Bohr's and Pauli's perspective within the context of crystallisation it will be shown that their attitudes are neither dogmatically conservative nor mysterious.  

\subsection{Bohr's Perspective}
Bohr belonged to the class of scientists who have a wide appreciation and understanding of the problems concerning the entire field to which they belong to, not counting a range of adjacent (i.e., pertaining to other sciences, particularly chemistry and biology) and non-adjacent problems (i.e., social, cultural, etc.).\footnote{ Insightful remarks regarding Bohr's intellectual preoccupations can be gleaned from Thomas Kuhn's, Aage Bohr's, and Leon Rosenfeld's interview with Margrethe Bohr, which can be found in the AIP Archives, see \cite{Bohr:1963}.} Through his visitors at the Copenhagen institute, through his correspondence, and through academic journals, Bohr liked to keep track of developments in physics and other sciences, even though such developments may not have been strictly relevant to his own research interests.\footnote{ John Hendry takes this to be the typical attitude of most physicists of the time, claiming that ``[even though they may have published only within restricted areas, most physicists took their perception of the problem complex from the wider field of their reading, correspondence and conversation" (\cite{Hendry:1984}, p. 6).} As regards his research interests, they were as varied as his intellectual reach. This notwithstanding, it is plausible to suggest that in the early '30s he was primarily interested in the structure of the atomic nucleus and the $\beta$-decay puzzles.\footnote{ Victor Weisskopf, for example, writes that Niels Bohr's work ``can be divided into four periods", where ``[t]he third period, 1930-1940, was devoted to the application of the new quantum concepts to electromagnetic phenomena and the exploration of the structure of the nucleus" (\cite{Weisskopf:1984}, pp. 584-585).} His interest in these problems did not stop him from suggesting solutions to related problems, though, as it will become apparent below, solutions to adjacent problems were inevitably assessed by their possible import to the main problems he was interested in.

Having heard from Gamow about the `$\pm$ difficulty', Bohr wrote to Dirac to ask for details, confessing that he is interested in this problem in connection to the $\beta$-decay problem, particularly ``as regards conservation of energy in $\beta$-ray disintegration and the interior of stars" (\cite{BCW:1986}, p. 547). In the same letter, he also suggested that just as in nuclear problems, ``an essential limitation of the principles of conservation of energy and momentum" (p. 548) might have to be faced, a departure from these principles might also be of use in solving the `$\pm$ difficulty'. Dirac, as we know, resisted this suggestion and in his letter to Bohr made it plain that he prefers ``to keep rigorous conservation of energy at all costs and would rather abandon even the concept of matter consisting of separate atoms and electrons than the conservation of energy".\footnote{ Dirac's remark is cited in \cite{Kragh:1990}, p. 90 and the full letter can be found in \cite{BCW:1986}, p. 548 - 550).} The conservation of energy principle ``had been in the history of physics an ever trustworthy guide" \cite{Darrigol:1988}, yet, as Helge Kragh writes, ``[in] the light of experimental anomalies and conceptual difficulties related to relativistic quantum mechanics [Bohr] foresaw a drastic revision of fundamental physics, with strict energy conservation being one of the necessary sacrifices" (\cite{Kragh:2017}, p. 126). We can already see that crystallisation invited the reevaluation of core assumptions such as the conservation of energy principle, yet against the background of crystallisation, Bohr's attitude is very far from being dogmatically conservative.

Bohr was coming to the problem of the negative energy solutions, via a perspective different from Dirac's and was more concerned with how Dirac's equation might bear on nuclear electrons and the problem of $\beta$-decay, than he was concerned with Dirac's hole theory as such. Bohr returned to the `$\pm$ difficulty' again and again, though on every occasion it was in connection to either the structure of the nucleus or the problem of $\beta$-decay. Bohr continues to `upset' the energy conservation throughout the early '30s and refers to this possibility again in his 1933 Solvay report, restating that he only ``advocated that one seriously consider the idea of a possible failure of the theorems of conservation of energy and momentum in connection with the continuous $\beta$-ray spectra" and in order to ``emphasise the total inadequacy of the classical conceptual edifice for treating this problem" (\cite{BCW:1996}, p. 191). It is clear from Bohr's remarks that the path of reevaluating core assumptions was not to be excluded in a period of crystallisation. 

Bohr's attitude to developments in connection to the `$\pm$ difficulty' did not change much throughout the early `30s since these developments were not presenting any viable solutions to either the problem of $\beta$-decay or to the structure of the nucleus. Dirac's prediction of the positron was largely ignored and it is plausible to suggest that insofar as Bohr is concerned his doubts regarding the positron are due to the fact that this new particle did not help resolve any of the problems he concerned himself with. His perspective, though different from Dirac's, was shaped by different concerns and a different, non-conservative attitude towards the reevaluation of background assumptions.\footnote{ A similar point is made by John Hendry, who notes that ``[t]he physicists concerned with a particular set of problems did not share the same influences and concerns; each one approached the current issues, both technical and philosophical, from a different framework" (\cite{Hendry:1984}, p. 4).} He continued to doubt Dirac's positron even after Blackett and Occhialini's report confirmed Anderson's findings and tentatively put forward a connection between the experimental results and Dirac's hole theory.\footnote{ In a footnote to a letter to Peierls, dated May 22, 1933, Pauli reports that ``[w]hen I saw him in March, he [Bohr] absolutely rejected the positive electron and thought that Blackett had just produced some `pathological photographs'" (\cite{BCW:1996}, p. 468).} In April 1933 he wrote the following to Oskar Klein: 
\begin{quote}
Regarding the positive electrons I cannot, however, quite share your enthusiasm. I am at least as yet very sceptical as regards the interpretation of Blackett's photographs, and am afraid that it will take a long time before we can have any certain knowledge about the existence or non-existence of the positive electrons. Nor as regards the applicability of Dirac's theory to this problem, I feel certain, or more correctly, I doubt it, at least for the moment (cited in \cite{Roque:1997}, p. 76).\footnote{ Rudolf Peierls recollects the same attitude, noting that ``[e]even after the discovery of the positron, which behaved exactly as Dirac had predicted a hole would behave, Bohr was cautious and urged that it should not be taken for granted that this was the right interpretation of the new particle" (\cite{Peierls:1985}, p. 58).}
\end{quote}

Bohr continued to be suspicious of the identification of the experimental particle with the Dirac particle throughout 1933. His attitude, far from being mysterious, was due to a number of factors that could not easily be put into a consistent picture. First, he expected that the transition from ``the physics of the atoms to that of nuclei may involve the sacrifice of some fundamental principle, namely energy conservation" (\cite{Enz:2002}, p. 214). Second, he initially believed that Blackett's and Occhialini's photographs were `pathological photographs', a belief which, as Roque points out, was not unfounded, since ``Ir\`ene  Curie and Fr\'ed\'eric Joliot had previously disregarded positively curved tracks as experimental artefacts, and Millikan had dismissed Anderson's original conclusions on the same grounds, prompting him to check on turbulence" (\cite{Roque:1997}, p. 85). Third, Blackett's and Occhialini's interpretation was tied to Dirac's hole theory, of which Bohr thought that it was ``not a theory that one can believe in"\footnote{ Bohr's remark as reported by Rosenfled in his interview with Oskar Klein, see \cite{Klein:1963}.} for a variety of reasons (i.e., formalism, measurement issues etc.). These different problems and assumptions shaped a perspective which led Bohr to resist the authentication of the positron.

\subsection{Pauli's Perspective}
The story of the positron cannot be told without an account of Pauli's reactions and views towards the events that shaped its authentication. Not only because ``everyone was eager to learn about Pauli's always forcefully and humorously expressed reactions to new discoveries and ideas, and his likes and dislikes of the prospects opened" (\cite{Bohr:1960}, p. 3), but, more importantly, because Pauli was an outspoken critic of Dirac's hole theory from the outset.\footnote{ Massimi, for example, notes that ``[s]ince its inception, Pauli had strongly opposed Dirac's hole picture as unphysical and counterintuitive, to the extent of welcoming Elsasser's hypothesis of the positron as an integral spin (or null spin) particle following Bose-statistics instead of Fermi-Dirac statistsics" (\cite{Massimi:2005}, p. 133).} He did not believe in Dirac's prediction and he maintained an ambivalent attitude towards the positron even after the discovery experiments were put forward, writing to Blackett that he doesn't ``believe on the Dirac-`holes', even if the positive electron exist" (sic)\footnote{ Pauli, Letter to Blackett, 19 April 1933, \cite{Pauli:1985}, p. 158.} and to Dirac that he doesn't believe in his ``perception of `holes', even if the existence of the `antielectron' is proved".\footnote{ Pauli, Letter to Dirac, 1 May 1933, \cite{Pauli:1985}, p. 159.} Pauli's attitude, however, is neither mystifying, nor dogmatically conservative or due to prejudice. On the contrary, understood against the background of crystallisation, Pauli's attitude is, like Bohr's, entirely reasonable and explicable, as will be shown in what follows. 

Pauli, like Bohr, liked to maintain a comprehensive view of the field. As one of his assistants recalls, [o]ne of the remarkable things about Pauli, really [...] is that he read everything. Everything!" (\cite{Uhlenbeck:1962}). It is not surprising then that the ``negative energy states of the Dirac equation [...] intrigued Pauli enormously" (\cite{Enz:2002}, p. 221). The `$\pm$ difficulty' did not, however, interest Pauli as much as other problems, such as the structure of the nucleus and the $\beta$-decay problem.\footnote{ Darrigol, for example, notes that during the early '30, Pauli ``indulged too much in nuclear electrons" (\cite{Darrigol:1988}, p. 247).} He applied himself to both of these problems by ``adhering to the already established \textit{laws} of energy conservation and of quantised angular momentum, making the necessary adjustments in the \textit{objects}, namely by proposing as working hypothesis the neutrino and the nuclear spin" (\cite{Enz:2002}, p. 228). Pauli's attitude towards the new realm of the nucleus, was not like Bohr's that of abandoning well established laws. He thought instead that ``[w]ith nuclear processes, even almost more important than the conservation laws of energy and momentum [...] are the conservation laws of all discreetly quantised quantities".\footnote{ Quoted in \cite{Enz:2002}, p. 228, full letter in \cite{Pauli:1985}, p. 184.} Pauli preferred to ``hold fast unconditionally to these assumptions and to pursue them in their consequences"\footnote{ Id. } before modifying them.\footnote{ According to Enz, ``Pauli was an unconditional believer in conservation laws" (\cite{Enz:2002}, p. 254).} In connection to the positron, the set of assumptions and problems of which Pauli's perspective consisted, played an important role in shaping his attitude towards its authentication. We can see these connections, i.e., strong adherence to entrenched background assumptions such as the conservation and symmetry laws as well as a commitment to the neutrino hypothesis, reiterated in his personal correspondence throughout 1933. In May, he writes the following to Peierls:

\begin{quote}
Furthermore, it is probable that the positive electron has integer (perhaps zero) spin and Bose statistics. This could, however, be deduced from the conservation laws (and from the empirical fact that the mass number of nuclei uniquely determines their statistics and the integer or half-integer character of their spin) only if one knew that an uneven number of them can be created in an elementary process. It would be desirable for experimenters to determine precisely the elementary process in which the positive electron is created. The idea that the positive electron obeys Bose statistics first came to Elsasser (albeit for different reasons) and I encouraged him to send a note on the matter to Nature. (The fact that this aspect is opposed to Dirac's hole theory only speaks \textit{for} it. For experience shows that positive and negative charge do not act exactly the same and I do not like at all moving this asymmetry into the initial state of the world, as Dirac does). (Letter to Peierls, 22 May 1933, translation by Alexander Blum in personal correspondence, full letter in \cite{Pauli:1985}, pp. 163-165.)
\end{quote}

Note that finding an explanation that would preserve both the conservation laws and the empirically established symmetry principle led Pauli to a differential interpretation of the discovery experiments. Being immersed in problems regarding the structure of the atomic nucleus, Pauli briefly entertained the Bose interpretation of the experimental particle.\footnote{ Pauli expresses his preference for a Bose-positron in a letter to Peierls dated 22 May 1933, as well as in a letter to Heisenberg dated 16 June 1933. A further letter to Heisenberg, dated 14 July 1933, marks the abandonment of the Bose-positron hypothesis, but is particularly instructive in explaining the the different problems and advantages associated with the Bose-positron. For the full letters see \cite{Pauli:1985}.} The Bose-positron hypothesis, whilst making the proton a composite particle, would also restore the elementarity of the neutron, a contentious issue at the time. However, more importantly, this hypothesis would help clear away some of the mysteries concerning $\beta-decay$: if the proton were to decay into a neutron and a Bose-positron the conservation of angular momentum would also be restored.\footnote{ Pauli's brief preference for the Bose-positron, along with the reasons and implications of this preference are discussed in \cite{Jensen:2000}, pp 172 -173, \cite{Massimi:2005}, pp. 133-134, see also fn. 78, and \cite{Darrigol:1988}, p. 257. Note that as Darrigol mentions, the idea of building the ``nuclei with neutrons and Bose positrons rather than with protons and neutrons" was still supported by Francis Perrin in September 1933, though he was convinced by others to abandon it. The point here is that such suggestions did not seem impossible in the context of rapid changes and discoveries characteristic of crystallisation.} It is plausible to suggest that Pauli's opposition to Dirac's positron was due to the fact that it did not shed light on either the nuclear puzzles or $\beta-decay$. To Pauli, the possibility of the experimental particle obeying Bose-Einstein statistics was appealing for yet another reason: it supported his own hypothesis about a new kind of particle that ``carries out'' the missing energy in $\beta-decay$: his neutrino might be a combination of a Bose-positron and an electron and this in turn would help maintain conservation of energy as well as solving the spin statistics anomalies. In this connection, he tells Peierls, that, 

\begin{quote}
When nuclear physics is concerned, I would again believe in the conservation laws - not only those of energy, momentum (and of course charge), but also of angular momentum and the symmetry characteristic (Bose or Fermi) of the whole system. As my early conjecture of the existence of ``neutrinos'' and its emission via beta-decay has been strengthened by the discovery of the positive electron. (Letter to Peierls, 22 May 1933, translation by Radin Dardashti in personal correspondence, full letter in \cite{Pauli:1985}, pp. 163-165).
\end{quote}

By mid June, Pauli still resisted the association between the discovered particle and the Dirac Particle, citing similar reasons as he gave to Peierls, in a letter to Heisenberg:

\begin{quote}
I do not believe in hole theory, because I wish to have an asymmetry between positive and negative charge in the laws of nature (it does not satisfy me to move the empirically observed asymmetry into the initial state of the world). Elsasser has voiced the hypothesis that the positive electrons might obey Bose statistics and have spin 0 or 1. I do not think that this can be excluded. Indeed, the fact that this would cause difficulties for Dirac's theory makes me like this hypothesis'. (Letter to Heisenberg, 16 June 1933, translation by Alexander Blum in personal correspondence, full letter in \cite{Pauli:1985}, p. 169.)
\end{quote}

By mid July 1933, Pauli gave up the Bose-positron idea, but he was still reluctant to equate the experimental particle with Dirac's theoretical prediction. It is worth noting that despite an intense interest in Dirac's equation and the hole theory, Pauli had misgivings about both from the beginning.\footnote{ Blum, in particular, stresses this point in connection to symmetry violations: ``The most prominent critic of the Dirac equation, and of hole theory in particular, was Pauli. He was unsatisfied with hole theory from the start" (\cite{Blum:2014}, p. 550).} As Blum notes, ``[w]hat bothered him most was the fact that it appeared to place the asymmetry between positive and negative charge in the initial conditions (i.e., the occupied negative energy states) rather than in the physical laws [...]" (\cite{Blum:2014}, p. 550). From Pauli' viewpoint, potential violations to conservation laws in $\beta$-decay also spoke against Dirac's positron, as did the formalism of the hole theory, for which Pauli used the ``deprecatory term `subtraction physics'" (\cite{Wentzel:1960}).

This succession of events suggests that Pauli's attitude, far from being due to prejudice against new particles, was instead shaped by carefully weighing his background assumptions against viable solutions to pressing conceptual and empirical problems in a period of crystallisation. Pauli did not believe in Dirac's anti-electron because the anti-electron was inherently tied to the hole theory, a theory that although he preoccupied himself with for a few years (even after the positron's authentication),\footnote{ Pauli, in collaboration with his assistant at the time, Victor Weisskopf, wrote an `anti-Dirac' paper \cite{Pauli:1934}, which was meant to avoid Dirac's `subtraction physics'.} he opposed it most of the time. And, he resisted Blackett's interpretation of the experimental results because it did not lead to viable solutions to the pressing conceptual and empirical problems he wished solved. Thus Pauli's reluctance, was not to one particle, but to two particles. 

Pauli's views exerted a certain influence in the community, since physicists ``always benefitted by Pauli's comments even if disagreement could temporarily prevail", but as Bohr notes, ``if he felt he had to change his views, he admitted it most gracefully, and accordingly it was a great comfort when new developments met with his approval".\footnote{ See Bohr's Foreword to the Pauli Memorial volume, \cite{Bohr:1960}, p. 3.} Pauli's change of heart regarding the positron, like his intermittent opposition to the positron throughout 1933, can be accounted for by understanding his perspective during crystallisation, in particular the problems he cared to solve and the theoretical assumptions he upheld. Bohr's remark only serves to provide additional support to the fact that Pauli's attitude was shaped by a concern with solutions to problems and not by prejudice or `personal animosity'. Pauli's biographer ponders ``[w]hy this personal animosity towards an abstract matter [i.e., Dirac's hole theory], one may ask, particularly since this theory was such a success in explaining the properties of the spinning electron? This is a difficult question to answer, even when considering that Pauli liked shocking language in his private statements" \cite{Enz:2002}, p. 295. As shown, Pauli's attitude was neither due to personal animosity nor exceedingly mysterious or difficult to account for.

By having examined the perspectives of Pauli and Bohr throughout 1933 we have gleaned some of the factors that prompted their reluctance to first, accept the positron as a new particle of matter; and second, to resist the equation of the Dirac particle with the experimental particle. Both Bohr and Pauli changed their attitude towards the positron by the end of 1933. Their changing perspectives, occasioned by certain developments prior to the 7th Solvay Conference, as well as by the discussions during the conference, have not yet been investigated. In the following section, the unfolding of the 7th Solvay Conference is therefore traced to glean further insights of Bohr's and Pauli's changing perspectives.  

\section{Authentication}
Abraham Pais claims that ``the positron theory as a serious discipline started in October 1933 with Dirac's address to the 7th Solvay Conference" (\cite{Pais:1989}, p. 98), whilst Bohr opens his Solvay remarks by praising ``[t]he wonderful confirmation which Dirac's theory of the electron has received through the discovery of the positron" (\cite{BCW:1996}, p. 183). Both suggest that the long authentication processes of the positron ended at the Solvay conference. Yet despite Bohr's remarks and Pais' statement, the discussions during the Solvay conference were still against the identification of the experiments with Dirac's theoretical prediction. It is more accurate to suggest that the necessary developments that led to the protracted equation of Dirac's prediction of the positron with the experimental positron were indeed set in motion at the Solvay Conference, so in a sense the discussions did mark the `the beginning of the positron as a serious discipline'. But it would be misleading, despite Bohr's pronouncement, to take the Solvay Conference as the locus of the positron's authentication. Bohr's and Pauli's interventions in connection to the positron, far from taking the experiments to confirm Dirac's prediction, suggested the contrary. They were both still sceptical of the identification of Dirac's theoretical prediction to the experimental particle, as will be shown below. It will be suggested, however, that developments with regard to the structure of the nucleus and the $\beta$-decay problem forced Bohr and Pauli to further rethink some of their previous assumptions and to reevaluate their stance towards the positron.

The 7th Solvay Conference on the ``Structure and properties of the atomic nucleus'' which took place between October 22nd -- October 29th 1933, in Bruxelles brought together some of the most influential physicists at the time, many of which were involved in one way or another in the positron story. As the title of the conference indicates, the main discussions revolved around the structure of the atomic nucleus with remarks on the positron occasionally looming in reports and in the discussions. In fact, Dirac's own report was a late addition to the conference's agenda, mainly due to an intervention of Pauli's.\footnote{ The quote from Pauli's letter to Paul Langevin is reproduced in \cite{Kragh:1990}, p. 146. See also \cite{Darrigol:1988} and \cite{Roque:1997} for reports on the 7th Solvay conference.} Whilst he was an ardent critic of the hole theory, Pauli thought that a ``report on the development of the hole theory and its relationship with the positive electron" (Pauli, letter to Langevin, cited in \cite{Kragh:1990}, p. 146), shorter and as a supplement to Heisenberg's report, would be most desirable and also useful for ``the general discussion of the theory of nuclei" (id., 146). The fact that Dirac was only invited due to Pauli's intervention, already suggests that little was hoped to be gained from Dirac's theory of holes in regard to the positron.

The conference begun with Langevin's opening remarks, followed by Cockcroft's report on the disintegration of elements by accelerated protons, Chadwick's report touching on the neutron discovery, and the Joliot-Curries' report on artificial radioactivity (\cite{Mehra:1975}, ch. 8, \cite{SolvayReports:1933}). Whilst Chadwick's report contributes towards `expelling' the electrons from the nucleus,\footnote{ Brown and Rechenberg \cite{Brown:1988} point out that, in places, Chadwick's report still suggests a composite neutron.} the Joliot-Curries' report offers further evidence of pair production, and thus of the existence of positrons.

Dirac's report at the Solvay conference was preceded by a favourable discussion by Blackett on the positive electron and its properties. Blackett's report drew mainly from the experimental data and he ``tactfully [...] discussed pair creation and annihilation in terms of the conservation of charge and energy alone" (\cite{Roque:1997}, p. 107). Whilst mostly favourable to Dirac, Blackett is fairly cautious in straightforwardly asserting that Dirac's theory has been conclusively confirmed by the experiments as can be seen from his remark:

\begin{quote}
Dirac's theory of the electron predicted particles with exactly the same properties, and so, the experiments provide some powerful support to the essential correctness of Dirac's theory. (author's translation from Solvay Report)\footnote{ La th\'eorie de l'\'electron de Dirac avait pr\'edit l'existence de particules ayant exactement ces m\^emes propri\'et\'es, de sorte que les r\'esultats de l'exp\'erience apportent un puissant appui en faveur de l'exactitude de la th\'eorie de Dirac quant \`a son essence (\cite{SolvayReports:1933}, p. 172).}
\end{quote}

Despite Blackett's confidence in Dirac's theory, a rather intense opposition to establishing a strong link between the hole theory and the experiments on the positive electron is apparent in the discussions. In particular, Blackett's arguments regarding the positive electron's spin, which would decisively show that the experimental particle is a Fermi-Dirac particle and hence should be equated with Dirac's prediction, were received with hostility by both Bohr  and Pauli as their comments clearly suggest: 

\begin{quote}
[Bohr] It is of utmost importance to learn, as Monsieur Blackett learns, to draw as many conclusions as possible from experiments on positive electrons, without having to appeal to Dirac's theory. I believe that the conclusion regarding charge is justified, but the conclusion regarding spin seems to me less certain. (author's translation from Solvay Report)\footnote{ ``Il est de la plus haute importance d'essayer, comme le fait M. Blackett, de tirer des conclusions aussi nombreuses que possible d'experi\'ences sur les \'electrons positifs, sans devoir recourir \`a la th\'eorie de Dirac. Je pense que la conclusion relative \`a la charge est juste, mais celle concernant le spin me semble moins certaine. En r\'ealit\'e, comme c'est la pr\'esence du champ nucl\'eaire qui permet au quantum de lumi\`ere incident de produire les deux particules, il n'st pas du tout exclu que le noyau prenne part au m\'ecanisme de la conservation du moment angulaire" (Bohr's comment to Blackett's report at the Solvay Conference (id.))}
\end{quote}

\begin{quote}
[Pauli] As opposed to what Monsieur Bohr thinks, I think that Monsieur Blackett's conclusion regarding the charge of the positive electron could not in any way be considered as more certain than his conclusion regarding its spin. (author's translation from Solvay Report)\footnote{ ``Contrairement \`a ce que pense M. Bohr, je suis d'avis que la conclusion de M. Blackett concernant la charge de l'\'electron positif ne saurait en aucune fa\c{c}on \^etre consid\'er\'ee comme plus certain que celle concernant son spin" (Pauli's comment to Blackett's report at the Solvay Conference (id.)).}
\end{quote}

It is important to note that ``at the Solvay congress of 1933 Bohr still resisted the evidence vindicating Dirac's anti-electron and contemplated the possibility that the positron's spin would differ from one half" (\cite{Darrigol:1988}, p. 256). In the discussions following Blackett's report on the positive electron, it is evident that Bohr and Pauli equally, pressed Blackett for details on how the  numerous conclusions with regard to the properties of the experimental particle could be derived \textit{without} appealing to Dirac's hole theory. Bohr, in particular, was still doubtful of the physical interpretation of spin, especially because such a property did not make much sense in the classical theory. For different reasons, Pauli resisted Blackett's arguments and in particular the suggestion that the experiments ought to be taken as a confirmation of Dirac's hole theory. 

Blackett's Solvay report failed to offer a definite theoretical identity to the positive electron. His proposed interpretation of the experimental results was not unquestioningly received. Besides Bohr and Pauli, Rutherford was also dissatisfied with the interpretation of the experiments. Whilst praising Blackett's experimental work, he expressed his concern with regard to the possibility that the experiments were influenced by Dirac's theory and he confessed that it would have been desirable to interpret the experiments without appeal to a preexisting theory.\footnote{ ``Il me semble qu'\`a certains r\'egards il est regrettable que nous ayons eu une th\'eorie de l'\'electron positif avant le d\'ebut des exp\'eriences. M. Blackett a fait son possible pour ne pas se laisser influencer par la th\'eorie, mais la fa\c{c}on d'envisager les r\'esultats doit in\'evitablement être influenc\'ee par la th\'eorie \'etait venue apr\`es l'\'etablissement exp\'erimental des faits" (Rutherford, comment to Blackett's report on the positive electron, in \cite{SolvayReports:1933}, p. 178).} More questions and comments followed, some leading to lengthy discussions and raising more suspicion with regard to the interpretation of the experimental results. In particular, a comment of Perrin's regarding the conservation of energy in the process of producing positrons stirred even more dissatisfaction from Bohr and Pauli. 

Blackett, however, like Ir\`ene  Curie, also reported on \textit{pair production}, with Blackett arguing that the production of a positive electron is not a nuclear phenomena, and hence pairs are created outside the nucleus (cf. \cite{SolvayReports:1933}, pp. 170 - 171). Blackett's hypothesis is reinforced by one of Lise Meitner's interventions in which she supports the idea that ``positrons, unlike beta electrons, did not emerge from the nucleus" (\cite{Guerra:2018}, p. 101),\footnote{ For Meitner's comment, see \cite{SolvayReports:1933}, pp. 175 - 176.} an idea embraced by Pauli, but opposed by Bohr. The idea that positron and electron pairs were produced outside the nucleus strengthened an already existing, though not predominant, suggestion according to which the electron was not a component of the nucleus. Most importantly, though, it lent further support to Dirac's prediction, as Dirac had already envisaged in his 1931 paper `the creation simultaneously of an electron and anti-electron'. Pair production, had yet other implications not only in connection to the structure of the nucleus, which would no longer contain electrons, but also in connection to $\beta-decay$. The missing energy in $\beta-decay$ could be attributed to the neutrino, a hypothesis favoured by Pauli, though not by Bohr. 

Dirac's own report, entitled ``Theory of the Positron", followed Blackett's report. Rather surprisingly, Dirac's report did not perpetuate the discussion about positive electrons and their properties. Most questions were in connection to internal problems of the hole theory, or with regard to certain calculations which Dirac had not yet made. One of Bohr's comments in particular, showcases once again, however, his hostile attitude towards the hole theory. In this comment, Bohr questions the very possibility that, at least some of the consequences of Dirac's hole theory will ever be vindicated by experiments.\footnote{ ``Je me demande si, en somme, une v\'erification exp\'erimentale de ces cons\'equences de la th\'eorie des lacunes doit être regard\'ee comme possible ou non", Bohr's comment to Dirac's Solvay report, in \cite{SolvayReports:1933}, p. 178).} 

Bohr's attitude changes dramatically towards the end of the discussions; he starts his general remarks by praising ``[t]he wonderful confirmation which Dirac's theory of the electron has received through the discovery of the positron" (\cite{BCW:1996}, p. 183). This change of heart, though likely prompted by the discussions, may have, in fact, occurred after the Solvay Conference as ``the published proceedings were in the making for at least 4 months after Solvay's guests left their rooms at the H\^{o}tel M\'etropole, even though the discussions following the reports were printed as if they were direct transcriptions, and preserved a casual tone" (\cite{Roque:1997}, p. 106). Moreover, ``[i]n February 1934 speakers were sent the galley proofs and gave the final touches to their interventions -- 4 months after the event, then as now not a negligible time gap in physics" (\cite{Roque:1997}, p. 107). Short of a miraculous conversion to the Dirac particle, Roqu\'e attributes Bohr's and Pauli's changing attitudes to ``theoretical and experimental practices employed in its manufacture, and then legitimated by discovery accounts that emphasised the coherence of Dirac, Anderson, and Blackett and Occhialini's work"  (\cite{Roque:1997}, p. 115). However, it is equally likely that it was not only ``manufacture and coherence efforts" that led to the authentication of the positron. Insofar as Bohr and Pauli are concerned negotiating the contribution of the positron to the problems they wished to solve, as well as the reevaluation of certain assumptions and of pending problems and solutions against the post Solvay Conference background, can be said to equally define their changing attitudes towards the positron. 

The Solvay Conference marked the abandonment of certain prior assumptions and welcomed the arrival of new ones. During the discussions of the Solvay Conference, it became clear that electrons cannot be confined within the nucleus and hence the assumption that the nucleus was constituted by a proton and an electron was belatedly abandoned. In light of experimental evidence attesting the existence of the positrons, as well as the discovery of Chadwick's neutron, the belief that there are only two elementary constituents of matter, was also abandoned. Some things stayed the same, such as the energy conservation principle, whilst others were enthusiastically received. For example, the ``newly discovered process of electron positron creation" (\cite{Darrigol:1988}, p. 261) led to fruitful new hypothesis regarding $\beta-decay$ and nuclear constitution whilst Pauli's neutrino idea gained ground and was later fruitfully exploited by Fermi.\footnote{ Soon after the Solvay Conference Fermi formulated a theory of $\beta-decay$ for which he sought publication in \textit{Nature}. His note was not accepted by\textit{Nature} due to its speculative character so it was instead published in \textit{La Ricerca Scientifica} towards the and of December 1933.} All these results of the reports and discussions of the Solvay Conference can be taken to support the suggestion that Bohr's and Pauli's conversion to the Dirac particle was not a miraculous conversion, but instead stemmed from a reevaluation of their relevant perspectives. Both Bohr and Pauli embraced the mechanism of pair production and with it they accepted the existence of Dirac's positron. Both also subscribed to the new structure of the nucleus without the electron as a permanent inhabitant, and welcomed the rapid proliferation of (some) new particles. Both scientists abandoned some of their theoretical commitments and adopted new ones. All these changes mark rational shifts in the epistemic attitudes of unprejudiced scientists, who resisted the authentication of the positron for good epistemic reasons.\footnote{ It is plausible to understand these shifts in terms of rational changes in the credences a scientist assigns to relevant theoretical hypotheses. I am indebted to an anonymous referee for pointing this out.} Whilst it may be cautious not to take the Solvay Conference as the very final step in the authentication of the positron, the reports and discussions certainly set in motion the relevant steps towards welcoming the positron as a new particle of matter. 

\section{Concluding Remarks}

The historical case study of the positron is both a blessing and a curse. The short, yet fascinating, early history of the positron has already lent itself to a number of philosophically rich analyses, in particular on the complex nature of experiments \cite{Roque:1997} and the structure of novel predictions \cite{Pashby:2012}. Here we have sought to glean further insights regarding the authentication of the positron. This is the blessing. The curse, as one might have guessed, resides, amongst other things, in the precarious nature of any final pronouncements as regards physicists' perspectives towards authentication during crystallisation. Barring final pronouncements, it was shown, however, that the physicists' attitudes towards the authentication of the positron are neither mystifying nor based upon dogmatic conservatism, but indeed that their attitudes are both reasonable and perfectly explicable if seen in the context of crystallisation. 

All scientific developments, not least the authentication of a new particle, take place within the historical context of a research tradition, and are viewed by the scientists of the day from a range of perspectives. When the research tradition in question is within a period of crystallisation, where basic concepts are being reevaluated, scientists with different perspectives may respond differentially to the same empirical knowledge, and this in turn may serve to make any process of authentication more fraught. Recognising this general pattern, which is well illustrated by the present discussion of the case of the positron, can help us understand the causes of certain types of scientific disagreements, and in doing so refine our narratives of scientific discovery in a manner more sensitive to both the peculiarities of the context and the diverse perspectives of the cast.  \\

\section*{Acknowledgements}

I would like to thank Michela Masimi, Karim Th\'ebault, Alexander Blum, Tom Pashby, Erik Curiel, Paul Teller, and Alain Th\'ebault for very helpful comments on early versions of this paper. Special thanks go to two anonymous reviewers whose comments helped me improve the paper significantly. My thanks also go to Michela Massimi, Karim Th\'ebault, Sean Gryb, James Ladyman, Don Ross, Tom Pashby, Stephan Hartman, and Nicos Stylianou for helpful conversations on the topic. Alexander Blum, Richard Dawid, and Radin Dardashti kindly translated bits of Pauli's letters from German for me, and C\`edric Paternotte and M\u{a}d\u{a}lina Guzun helped me with some translations from French. Further thanks go to audiences at Edinburgh, Exeter, Nottingham, and Bristol for very helpful comments on earlier versions of this paper. This article feeds into a larger project that has received funding from the European Research Council (ERC) under the European Union's Horizon 2020 research and innovation programme (grant agreement European Consolidator Grant H2020-ERC-2014-CoG 647272 Perspectival Realism. Science, Knowledge, and Truth from a Human Vantage Point).

\bibliography{References}

\begin{thebibliography}{10}
\expandafter\ifx\csname url\endcsname\relax
  \def\url#1{\texttt{#1}}\fi
\expandafter\ifx\csname urlprefix\endcsname\relax\def\urlprefix{URL }\fi
\expandafter\ifx\csname href\endcsname\relax
  \def\href#1#2{#2} \def\path#1{#1}\fi

\bibitem{Hanson:1961}
N.~R. Hanson, Discovering the positron (i), British Journal for the Philosophy
  of Science (1961) 194--214.

\bibitem{Darrigol:1988}
O.~Darrigol, The quantum electrodynamical analogy in early nuclear theory or
  the roots of {Y}ukawa's theory, Revue d'histoire des sciences (1988)
  225--297.

\bibitem{Kragh:1990}
H.~Kragh, Dirac: a scientific biography, Cambridge University Press, 1990.

\bibitem{Roque:1997}
X.~Roqu{\'e}, The manufacture of the positron, Studies in History and
  Philosophy of Science Part B: Studies in History and Philosophy of Modern
  Physics 28~(1) (1997) 73--129.

\bibitem{Pashby:2012}
T.~Pashby, Dirac's prediction of the positron: A case study for the current
  realism debate, in: Perspectives on Science, MIT Press, 2012, pp. 440--475.

\bibitem{Leone:2012}
M.~Leone, N.~Robotti, An uninvited guest: The positron in early 1930s physics,
  American Journal of Physics 80~(6) (2012) 534--541.

\bibitem{Laudan:1977}
L.~Laudan, Progress and its Problems: Towards a theory of scientific growth,
  1978th Edition, University of California Press of California Press, 1977.

\bibitem{Kao:2015}
M.~Kao, Unification and the quantum hypothesis in 1900 - 1913, Philosophy of
  Science 82~(5) (2015) 1200--1210.

\bibitem{Massimi:2016a}
M.~Massimi, Four kinds of perspectival truth, Philosophy and Phenomenological
  Research 96~(2) (2016) 342 -- 359.

\bibitem{Massimi:2018a}
M.~Massimi, Perspectivism, in: J.~Saatsi (Ed.), The Routledge Handbook of
  Scientific Realism, Routledge, 2018, Ch.~13, pp. 164 -- 175.

\bibitem{Giere:2006}
R.~N. Giere, Perspectival pluralism, in: C.~K.~W. Stephen H.~Kellert, Helen
  E.~Longino (Ed.), Scientific Pluralism, Minneapolis: University of Minnesota
  Press, 2006, pp. 26--41.

\bibitem{Giere:2010}
R.~N. Giere, Scientific perspectivism, University of Chicago Press, 2010.

\bibitem{DeCruz:2013}
H.~De~Cruz, J.~De~Smedt, The value of epistemic disagreement in scientific
  practice. the case of {Homo} floresiensis, Studies in History and Philosophy
  of Science Part A Vol: 44~(Issue: 2).

\bibitem{Chadwick:1932}
J.~Chadwick, Possible existence of a neutron, Nature 129~(3252) (1932) 312.

\bibitem{Stuewer:2012}
R.~H. Stuewer, The nuclear electron hypothesis, in: Otto Hahn and the rise of
  nuclear physics, Springer Science, 2012, pp. 19 -- 67.

\bibitem{Bromberg:1971}
J.~Bromberg, The impact of the neutron: Bohr and {H}eisenberg, Historical
  Studies in the Physical Sciences 3 (1971) 307--341.

\bibitem{Stuewer:2018}
R.~H. Stuewer, The Age of Innocence: Nuclear Physics Between the First and
  Second World Wars, Oxford University Press, 2018.

\bibitem{Pais:1991}
A.~Pais, Niels {B}ohr's Times, in physics, Philosophy, and Polity, Clarendon
  Press Oxford, 1991.

\bibitem{Guerra:2014}
F.~Guerra, M.~Leone, N.~Robotti, When energy conservation seems to fail: the
  prediction of the neutrino, Science and Education 23~(6) (2014) 1339 -- 1359.

\bibitem{Jensen:2000}
C.~Jensen, Controversy and Consensus: Nuclear Beta Decay 1911 -- 1934, Vol.~24,
  Springer Basel AG, 2000.

\bibitem{Brown:1988}
L.~M. Brown, H.~Rechenberg, Nuclear structure and beta decay (1932--1933),
  American Journal of Physics 56~(11) (1988) 982--988.

\bibitem{Kragh:2017}
H.~Kragh, `{L}et the stars shine in peace!' {N}iels {B}ohr and stellar energy,
  1929 - 1934, Annals of science 74~(2) (2017) 126--148.

\bibitem{Guerra:2018}
F.~Guerra, N.~Robotti, The Lost Notebook of Enrico Fermi, Springer, 2018.

\bibitem{Navarro:2004}
J.~Navarro, New entities, old paradigms: Elementary particles in the 1930s,
  Llull: Revista de la Sociedad Espa{\~n}ola de Historia de las Ciencias y de
  las T{\'e}cnicas 27~(59) (2004) 435--464.

\bibitem{Dirac:1928}
P.~A.~M. Dirac, The quantum theory of the electron, in: Proceedings of the
  Royal Society of London A: Mathematical, Physical and Engineering Sciences,
  Vol. 117, The Royal Society, 1928, pp. 610--624.

\bibitem{Dirac:1980}
P.~A.~M. Dirac, The origin of quantum field theory, in: L.~M. Brown,
  L.~Hoddeson (Eds.), The Birth of Particle Physics, Cambridge University
  Press, 1983, pp. 39 -- 55.

\bibitem{Dirac:1973}
P.~A.~M. Dirac, Development of the physicist's conception of nature, in:
  J.~Mehra (Ed.), The Physicist's Conception of Nature, D. Reidel Publishing
  Company, 1973.

\bibitem{Dirac:1962}
W.~E. Kuhn, Thomas~S., P.~A.~M. Dirac, Interview of {P}. {A}. {M}. {D}irac by
  {T}homas {S}. {K}uhn and {E}ugene {W}igner, Niels Bohr Library \& Archives,
  American Institute of Physics, College Park, MD USA
  www.aip.org/history-programs/niels-bohr-library/oral-histories/4575-1.

\bibitem{Moyer:1981a}
D.~F. Moyer, Origins of {D}irac's electron, 1925--1928, American journal of
  physics 49~(10) (1981) 944--949.

\bibitem{Kragh:1981}
H.~Kragh, The genesis of {D}irac's relativistic theory of electrons, Archive
  for History of Exact Sciences 24~(1) (1981) 31--67.

\bibitem{Massimi:2005}
M.~Massimi, Pauli's exclusion principle: The origin and validation of a
  scientific principle, Cambridge University Press, 2005.

\bibitem{Blum:2014}
A.~Blum, From the necessary to the possible: the genesis of the spin-statistics
  theorem, The European Physical Journal H 39~(5) (2014) 543--574.

\bibitem{Pais:1989}
A.~Pais, On the {D}irac theory of the electron (1930 - 1936),
  https://link.springer.com/content/pdf/bfm

\bibitem{Moyer:1981b}
D.~F. Moyer, Evaluations of {D}irac's electron, 1928--1932, American journal of
  physics 49~(11) (1981) 1055--1062.

\bibitem{BCW:1986}
N.~Bohr, Nuclear Physics (1929-1952), Vol.~9 of Niels Bohr Collected Works,
  North-Holland Physics Publishing, 1986.

\bibitem{Dirac:1930a}
P.~A.~M. Dirac, A theory of electrons and protons, in: Proceedings of the Royal
  Society of London A: Mathematical, Physical and Engineering Sciences, Vol.
  126, The Royal Society, 1930, pp. 360--365.

\bibitem{Weyl:1950}
H.~Weyl, The theory of groups and quantum mechanics, Courier Corporation, 1950.

\bibitem{Pais:1987}
A.~Pais, Playing with equations, the {D}irac way, in: B.~N. Kursunoglu, E.~P.
  Wigner (Eds.), Reminiscences about a Great Physicist: Paul Adrien Maurice
  Dirac, Cambridge University Press, 1987, pp. 93 -- 116.

\bibitem{Dirac:1930b}
P.~A.~M. Dirac, The proton, Nature 126 (1930) 605--606.

\bibitem{Dirac:1931}
P.~A.~M. Dirac, Quantised singularities in the electromagnetic field, in:
  Proceedings of the Royal Society of London A: Mathematical, Physical and
  Engineering Sciences, Vol. 133, The Royal Society, 1931, pp. 60--72.

\bibitem{Anderson:1961}
C.~D. Anderson, Early work on the positron and muon, American Journal of
  Physics 29~(12) (1961) 825--830.

\bibitem{Pais:1998}
A.~Pais, Paul {D}irac: aspects of his life and work, in: P.~Goddard (Ed.), Paul
  {D}irac: the man and his work, Cambridge University Press, 1998.

\bibitem{Jacob:1998}
M.~Jacob, Antimatter, in: P.~Goddard (Ed.), Paul Dirac: the man and his work,
  Cambridge University Press, 1998.

\bibitem{Griffiths:2004}
D.~Griffiths, Introduction to elementary particles, WILEY-VCH, 2004.

\bibitem{Heilbron:1962}
J.~L. Heilbron, Interview of {P}. {M}. {S}. {B}lackett by {J}ohn {L}.
  {H}eilbron, Niels Bohr Library \& Archives, American Institute of Physics,
  College Park, MD USA
  www.aip.org/history-programs/niels-bohr-library/oral-histories/4508.

\bibitem{Chadwick:1969}
C.~Weiner, J.~Chadwick, Interview of {J}ames {C}hadwick by {C}harles {W}einer,
  Niels Bohr Library \& Archives, American Institute of Physics, College Park,
  MD USA www.aip.org/history-programs/niels-bohr-library/oral-histories/3974-1.

\bibitem{Klein:1963}
J.~L. Heilbron, L.~Rosenfeld, O.~Klein, Interview of {O}skar {K}lein by {J}.
  {L}. {H}eilbron and {L}. {R}osenfeld, Niels Bohr Library \& Archives,
  American Institute of Physics, College Park, MD USA
  www.aip.org/history-programs/niels-bohr-library/oral-histories/4709-4.

\bibitem{Kojevnikov:1993}
A.~B. Kojevnikov, Paul {D}irac and {I}gor {T}amm Correspondence: Part 1 1928 -
  1933, http://cds.cern.ch/record/258359/files/P00020744.pdf, 1993.

\bibitem{Bohr:1963}
A.~B. Margarethe~Bohr, Thomas S.~Kuhn, L.~Rosenfeld, Interview of {M}argarethe
  {B}ohr by {T}homas {S}. {K}uhn, {A}age {B}ohr, and {L}eon {R}osenfeld, Niels
  Bohr Library \& Archives, American Institute of Physics, College Park, MD USA
  www.aip.org/history-programs/niels-bohr-library/oral-histories/4514-1.

\bibitem{Hendry:1984}
J.~Hendry, The Creation of Quantum Mechanics and the Bohr-Pauli Dialogue,
  Vol.~14 of Studies in the History of Modern Science, D. Reidel Publishing
  Company, 1984.

\bibitem{Weisskopf:1984}
V.~Weisskopf, Niels {B}ohr, the {Q}uantum, and the {W}orld, Social Research
  51~(3) (1984) 583 -- 608.

\bibitem{BCW:1996}
N.~Bohr, Foundations of Quantum Mechanics II (1933 -1958), Vol.~7, Elsevier
  Science B.V., 1996.

\bibitem{Peierls:1985}
R.~Peierls, Bird of Passage. Recollections of a Physicist, Princeton University
  Press, 1985.

\bibitem{Enz:2002}
C.~P. Enz, No Time to be Brief: A scientific biography of Wolfgang Pauli,
  Oxford University Press, 2002 (Reprinted 2013).

\bibitem{Bohr:1960}
N.~Bohr, Foreword, in: M.~Fierz, V.~F. Weisskopf (Eds.), Theoretical Physics in
  the Twentieth Century: A Memorial Volume to Wolfgang Pauli, Interscience
  Publishers Inc., New York, 1960, pp. 1 -- 4.

\bibitem{Pauli:1985}
W.~Pauli, Scientific Correspondence with Bohr, Einstein, Heisenberg a.o. Volume
  II: 1930-1939, Springer-Verlag, 1985.

\bibitem{Uhlenbeck:1962}
G.~Uhlenbeck, T.~S. Kuhn, Interview of {G}eorge {U}hlenbeck by {T}homas {S}.
  {K}uhn, Niels Bohr Library \& Archives, American Institute of Physics,
  College Park, MD USA
  www.aip.org/history-programs/niels-bohr-library/oral-histories/4922-1.

\bibitem{Wentzel:1960}
G.~Wentzel, Quantum theory of fields (until 1947), in: M.~Fierz, V.~F.
  Weisskopf (Eds.), Theoretical Physics in the Twentieth Century: A Memorial
  Volume to Wolfgang Pauli, Interscience Publishers Inc., New York, 1960, pp.
  48 -- 77.

\bibitem{Pauli:1934}
W.~Pauli, V.~Weisskopf, On quantization of the scalar relativistic wave
  equation, Helv. Phys. Acta 7 (1934) 709--731.

\bibitem{Mehra:1975}
J.~Mehra, The Solvay Conferences on Physics, D. Reidel Publishing Company,
  1975.

\bibitem{SolvayReports:1933}
Gauthier-Villars (Ed.), Structure et propri\'et\'es des noyaux atomiques:
  rapports et discussions du septi\'eme Conseil de physique tenu \'a Bruxelles
  du 22 au 29 octobre 1933, sous les auspices de l'Institut international de
  physique Solvay, Commission Administrative de l'Institut Solvay, 1935.

\end{thebibliography}

\end{document}